# Using Bluetooth Low Energy in smartphones to map social networks

Samuel Townsend [a], Mark E. Larsen [a], Tjeerd W. Boonstra [a] and Helen Christensen [a]

[a] Black Dog Institute, University of New South Wales, Sydney, Australia

Email: s.townsend@unsw.edu.au, mark.larsen@blackdog.org.au, t.boonstra@unsw.edu.au, h.christensen@blackdog.org.au

**Abstract**

Social networks have an important role in an individual's health, with the propagation of health-related features through a network, and correlations between network structures and symptomatology. Using Bluetooth-enabled smartphones to measure social connectivity is an alternative to traditional paper-based data collection; however studies employing this technology have been restricted to limited sets of homogenous handsets. We investigated the feasibility of using the Bluetooth Low Energy (BLE) protocol, present on users' own smartphones, to measure social connectivity. A custom application was designed for Android and iOS handsets. The app was configured to simultaneously broadcast via BLE and perform periodic discovery scans for other nearby devices. The app was installed on two Android handsets and two iOS handsets, and each combination of devices was tested in the foreground, background and locked states. Connectivity was successfully measured in all test cases, except between two iOS devices when both were in a locked state with their screens off. As smartphones are in a locked state for the majority of a day, this severely limits the ability to measure social connectivity on users' own smartphones. It is not currently feasible to use Bluetooth Low Energy to map social networks, due to the inability of iOS devices to detect another iOS device when both are in a locked state. While the technology was successfully implemented on Android devices, this represents a smaller market share of partially or fully compatible devices.

**Introduction**

Social networks have an important role in an individual's health and wellbeing, and features such as obesity (1), smoking status and happiness (2) have been shown to propagate through an individual's network. The topology of social networks has also been shown to relate to mental health, such as the clustering of individuals with depressive symptoms (3) and differing degrees of connectivity (4). Changes in social connectivity can also be indicative of change in mental health state and significant events, such as suicidal behaviour (5, 6).

Traditional techniques for mapping social network structures involve using name generators to ask people to self-report their degree of connectivity with a range of identified individuals. Smartphones with Bluetooth have been used to accurately infer friendship networks (7), indicating the potential for mobile apps to measure social networks (8). However, previous studies using Bluetooth technology in smartphones have relied on providing study-specific handsets, which limits the scalability of such data collection processes. We propose using users' own handsets as a means of facilitating studies into social networks.

Although Bluetooth has been shown to be a suitable technology for inferring social networks information, it is not officially supported on iOS devices. The newer Bluetooth Low Energy (BLE) protocol is, however, supported on iOS devices as well as newer Android handsets. BLE is used in iBeacon technology (9) to detect when handsets are in proximity to a particular device, which suggests it is a suitable platform for detecting other nearby users as a basis for mapping social networks.

Here we investigate the capabilities of BLE technology implemented in a custom application designed for Android and iOS devices. Implementing this technology on users' own handsets will enable mapping social networks on much larger scale.

**Methods**

A custom application was designed for both Android and iOS devices which utilises the Bluetooth Low Energy protocol to determine the user's social network. This application was targeted at Android devices running KitKat (version 4.4) and above (52% of Android devices as of June 2015 (10)) and Apple devices running iOS 7 and above (98% of iOS devices as of June 2015 (11)).

The app was configured to simultaneously advertise the handset's presence via BLE, and to perform a periodic discovery scan for other nearby devices. Following previous work, the handset's Media Access Control (MAC) address was initially used to uniquely identify each handset and user. A MAC address is a unique identifier which is used by a device's network interface, such as WiFi and Bluetooth chips present in a smartphone, in order to communicate and direct the flow of data on a network segment. These identifiers are usually written into the hardware itself by the manufacturer, and are generally guaranteed to be unique between devices.

However, as of iOS 8, these unique and consistent MAC addresses have been replaced by random, constantly changing addresses. This is a privacy-maintaining feature (12) to restrict passive tracking of user habits and behaviours by third parties without consent. This feature is not present in current Android devices, which provide access to the device's unique MAC address. To allow social connections to be mapped in the absence of a consistent MAC address, a custom BLE service was implemented. This allowed detection only by other handsets that shared the pre-defined service identifier. Each handset generated a unique identifier which was advertised as a characteristic within the shared service. This identifier could then be shared with other handsets with the app installed, allowing all users to be correctly identified.

To test the ability to map social networks using BLE, the custom app was installed on a Samsung Galaxy S4 mini, a Samsung Galaxy S5, iPhone 5s, and an iPad mini. Initial testing involved using each handset in three states: foreground (when the app is shown to the user in the foreground, and the screen is active), background (when the app is running in running the background, the home screen is active, and the screen is on), and locked (when the app is running in the background, the home screen is active, and the screen is off). A series of tests were performed to determine if each of the devices could detect each other in the three different app states. For initial testing, only two devices were used in order to avoid the possibility of extra devices impacting on the ability to correctly detect the other device. This meant the testing of three combinations (Android to Android, Android to iOS, and iOS to iOS), with each pair being tested in one of the three states outlined above. For the Android to iOS combination, the test was repeated with each combination of device to ensure consistent results were obtained.

## Results

The results of testing the devices in the various configurations are shown in Table 1. Each device could detect each other in all configurations, except one: two iOS devices could not detect each other when they were both in the locked state. This meant that while the iOS users would be able to record their interactions with other users with Android devices, they would not detect an interaction with other users with an iOS device that is in a locked state.

**Table 1: Results of tests to determine if handsets could detect each other.**

| Device Configuration | | Android | | | iOS | | |
|---|---|---|---|---|---|---|---|
| | | FB | BG | L | FG | BG | L |
| **Android** | FG | Pass | Pass | Pass | Pass | Pass | Pass |
| | BG | Pass | Pass | Pass | Pass | Pass | Pass |
| | L | Pass | Pass | Pass | Pass | Pass | Pass |
| **iOS** | FG | Pass | Pass | Pass | Pass | Pass | Pass |
| | BG | Pass | Pass | Pass | Pass | Pass | Pass |
| | L | Pass | Pass | Pass | Pass | Pass | Fail |

FB = app in foreground state, BG = app in background state, L = app in locked state. Pass = handsets detected each other, Fail = handsets failed to detect each other.

While this may seem to impact only a small fraction of use cases, in fact this is the largest use case. A recent Nielsen study measured smartphone usage at an average of 37 hours per month (13), equating to a period of 1hr and 15 minutes per day for which a user's phone could be expected to be unlocked. Assuming a sleep duration of eight hours per day, during which time no connectivity would be detected, this means an iOS handset would be unable to detect other locked iOS devices for 92% of the waking day. Therefore the probability of any two iOS devices failing to detect each other during this 16-hour period is 85%. The ability to detect devices during this time is therefore critical for the implementation of this technology.

The reason for the inability to detect locked iOS devices is not clear, but appears to be related to the operating system partially removing advertising data from the BLE packet when the device is in a locked state. This is unlikely to be the sole reason, however, as Android devices and unlocked iOS devices can still detect a locked iOS device broadcasting this reduced packet.

## Conclusions

Using Bluetooth Low Energy to map social networks appears to be unviable at this time, due to inability of iOS devices to detect one another when both devices are in a locked state with their screens off. However, iOS devices in locked state advertised and scanned periodically, as they could still detect and be detected by Android devices. While the app functioned successfully on Android devices, the current market share of handsets which are partially compatible (version 4.4 and above) has not reached saturation, with fewer being fully compatible (version 5.0 and above) with the BLE technology. Future updates of the operating system may resolve the issue with iOS devices, during which time the number of compatible Android devices will increase.

## Acknowledgements

This project was supported by the NHMRC Centre of Research Excellence in Suicide Prevention APP1042580, NHMRC John Cade Fellowship APP1056964 and the Young and Well Cooperative Research Centre.

# References

1. Christakis NA, Fowler JH. The spread of obesity in a large social network over 32 years. The New England journal of medicine. 2007;357(4):370-9.

2. Fowler JH, Christakis NA. Dynamic spread of happiness in a large social network: longitudinal analysis over 20 years in the Framingham Heart Study. Bmj. 2008;337:a2338.

3. Rosenquist JN, Fowler JH, Christakis NA. Social network determinants of depression. Molecular psychiatry. 2011;16(3):273-81.

4. Pachucki MC, Ozer EJ, Barrat A, Cattuto C. Mental health and social networks in early adolescence: a dynamic study of objectively-measured social interaction behaviors. Social science & medicine. 2015;125:40-50.

5. Van Orden KA, Witte TK, Cukrowicz KC, Braithwaite SR, Selby EA, Joiner TE, Jr. The interpersonal theory of suicide. Psychological review. 2010;117(2):575-600.

6. Larsen ME, Cummins N, Boonstra T, O'Dea B, Tighe J, Nicholas J, et al., editors. The use of technology in suicide prevention. Conference of the IEEE Engineering in Medicine and Biology Society; 2015.

7. Eagle N, Pentland AS, Lazer D. Inferring friendship network structure by using mobile phone data. Proceedings of the National Academy of Sciences of the United States of America. 2009;106(36):15274-8.

8. Chronis I, Madan A, Pentland A, editors. SocialCircuits: the art of using mobile phones for modeling personal interactions. ICMI-MLMI '09 Workshop on Multimodal Sensor-Based Systems and Mobile Phones for Social Computing 2009 2009.

9. Apple. iOS: Understanding iBeacon  [6 July 2015]. Available from: https://support.apple.com/en-us/HT202880.

10. Google. Android Platform Versions Dashboard  [6 July 2015]. Available from: https://developer.android.com/about/dashboards/.

11. Apple. App Store Distribution Support  [6 July 2015]. Available from: https://developer.apple.com/support/app-store/.

12. IEEE 802 EC Privacy Recommendation Study Group  [6 July 2015]. Available from: http://www.ieee802.org/PrivRecsg/.

13. The Nielsen Company. How smartphones area changing consumers' daily routines around the globe 2015 [6 July 2015]. Available from: http://www.nielsen.com/us/en/insights/news/2014/how-smartphones-are-changing-consumers-daily-routines-around-the-globe.html.